\documentclass[fleqn,10pt]{wlscirep}
\usepackage[utf8]{inputenc}
\usepackage[T1]{fontenc}
\title{Fano-like spectral features in nanodiamond solutions for biometric applications}

\author[1,2*]{Graciana Puentes}
\affil[1]{Departamento de Fsica, Facultad de Ciencias Exactas y Naturales, Universidad de
Buenos Aires, Ciudad Universitaria, 1428 Buenos Aires, Argentina
}
\affil[2]{CONICET-Universidad de Buenos Aires, Instituto de Fsica de Buenos Aires
(IFIBA), Ciudad Universitaria, 1428 Buenos Aires, Argentina}

\affil[*]{gpuentes@df.uba.ar}

\begin{abstract}
Fano resonance is a unique feature of interacting quantum systems, exhibiting  resonance
shapes distinctively different from conventional symmetric resonance curves. Recently,  Fano resonances have been found in plasmonic nanoparticles, photonic crystals, and electromagnetic metamaterials. Here we report Fano-like photoluminiscence curves in nanodiamond solutions as a result of incoherent combination of two or more scattering and fluorescence processes. We argue that, analogously to Fano resonances, the  steep asymmetric dispersion of the photoluminiscence profile in nanodiamond solutions, in  combination with biologically-compatible spectral features characterizing nanodiamond fluorescence, can find promising biometric applications in several areas such as bio-sensors, bio-switches and bio-filters.
\end{abstract}
\begin{document}

\flushbottom
\maketitle
%
%
\thispagestyle{empty}

\section*{Introduction}

For many years, the fundamental lineshape of a resonance was regarded as the Lorentzian form. However, it is well known that such  spectral feature can be modified by the presence of independent resonances of different physical origins, where the lineshape is simply the sum of the intensities of the individual resonances that contribute to it \cite{1,2,3}. In contrast to a Lorentzian resonance, and as a result of the interference between different scattering processes, the Fano resonance exhibits a distinctly asymmetric shape with the following functional form:

\begin{center}
\begin{equation}
    I \propto \frac{F \gamma + \omega - \omega_0}{(\omega - \omega_0)^2+\gamma^2}
\end{equation}
\end{center}

where $\omega_0$  and $\gamma$ denote the central frequency and width of the resonance, respectively, $F$ is the so-called Fano parameter, which describes the degree of asymmetry. In Figure 1 (a), numerical simulations of the intensity profile associated with Fano resonances, for a central frequency $\omega_0=5.45 \times 10^{-14}$ Hz, and a linewidth $\gamma=1 \times 10^{15}$ Hz are depicted. The microscopic origin of the Fano resonance arises from the constructive and destructive interference of a narrow discrete resonance with a broad spectral line or continuum. Subsequent to its discovery, there have been a great number of studies devoted to Fano resonances in various quantum systems, such as quantum dots, nanowires and tunnel junctions \cite{4,5,6,7,8,9,10, 11, 12, 13 ,14,27,28,29}. \\

Nanodiamond (ND) particles are currently utilized as diagnostic tools for imaging in biomedicine, for cryptography and quantum information processing, and as single-spin sensors in nanomagnetometry \cite{15,16,17,18,19,20,21,22,23,24,25,26}. Most of these applications are based on the unique optical and magnetic properties associated with point defects in diamond. Due to the presence of wide-bandgap, NDs often contain atomic defects or impurities, some of which are highly luminescent making them useful as fluorescent agents. NDs are highly biocompatible due to their inherently low toxicity, among other unique features.

\begin{figure}[h]
\centering
\includegraphics[width=0.8\linewidth]{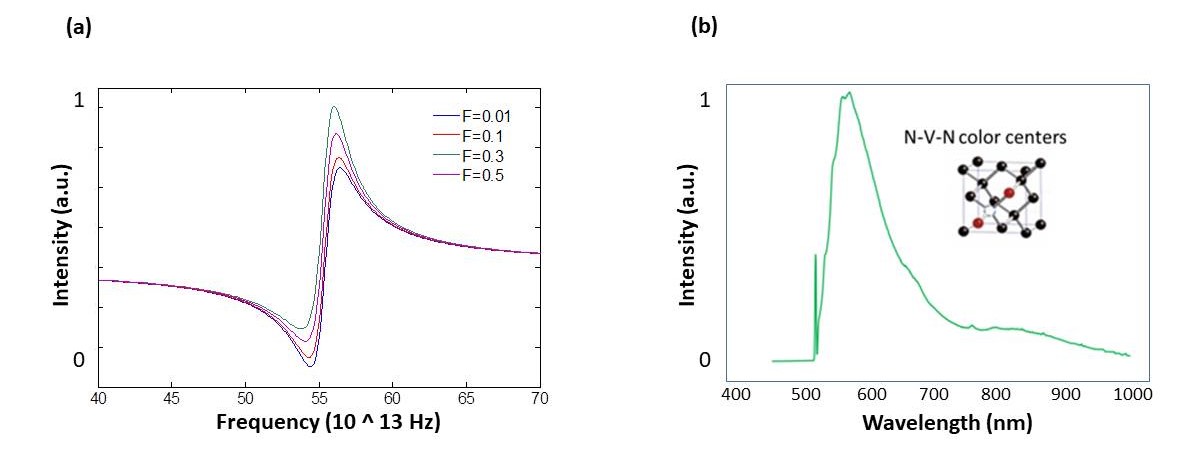}
\caption{(a) Numerical simulation of intensity characterizing Fano resonances (Eq. 1), for $\gamma=1 \times 10^{15}$ Hz and $\omega_0=5.54 \times 10^{14}$ Hz, different curves correspond to different Fano asymmetry parameters in the range $F=0.01-0.5$. (b) Typical fluorescence photo-luminiscence emitted by N-V-N centers dispersed in DI water. The sharp asymmetric shape of the spectral trace is apparent, revealing the Fano-like features characterizing N-V-N nanodiamond solutions (see Ref. [30]).}
\label{fig:stream}
\end{figure}

\begin{figure}[ht]
\centering
\includegraphics[width=0.9\linewidth]{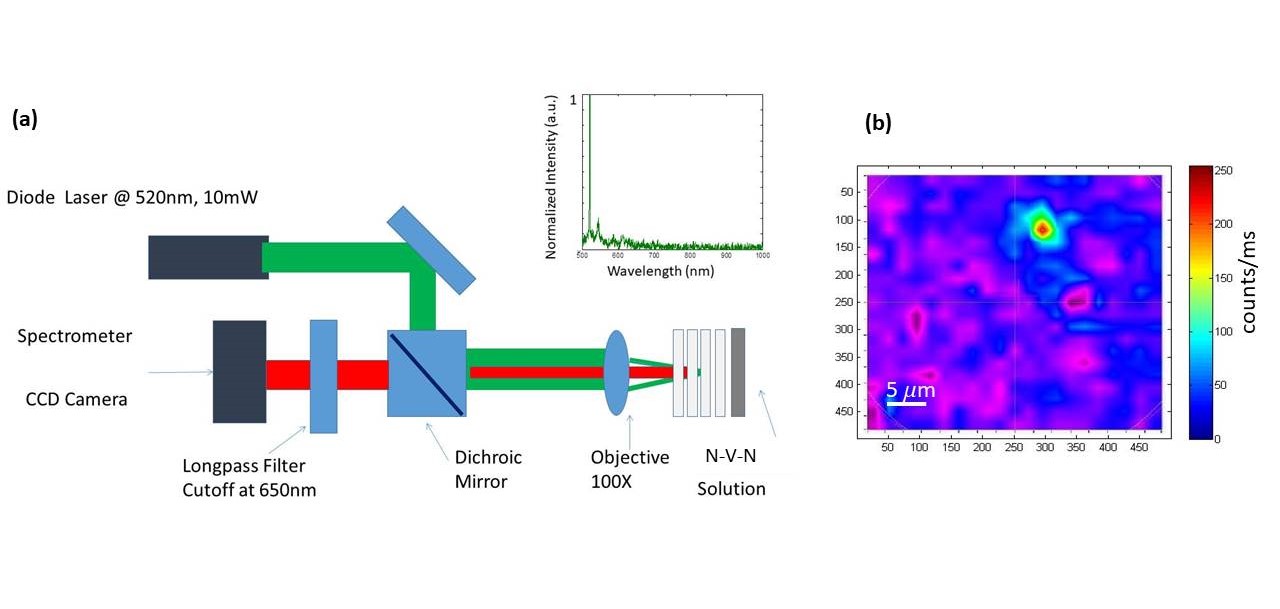}
\caption{(a) Schematic of fluorescence measurement setup. The spectrometer (Thorlabs CCS series) used for recording spectral traces was replaced by a CCD camera (Thorlabs WFS series) to perform confocal images. Inset: Normalized spectral trace caracterizing excitation laser (a.u.). (b)  NV confocal image using a CCD camera (Thorlabs WFS series). Intensity scale ranging from 0-250 cts/ms.}
\label{fig:stream}
\end{figure}

\begin{figure}[h!]
\centering
\includegraphics[width=0.7\linewidth]{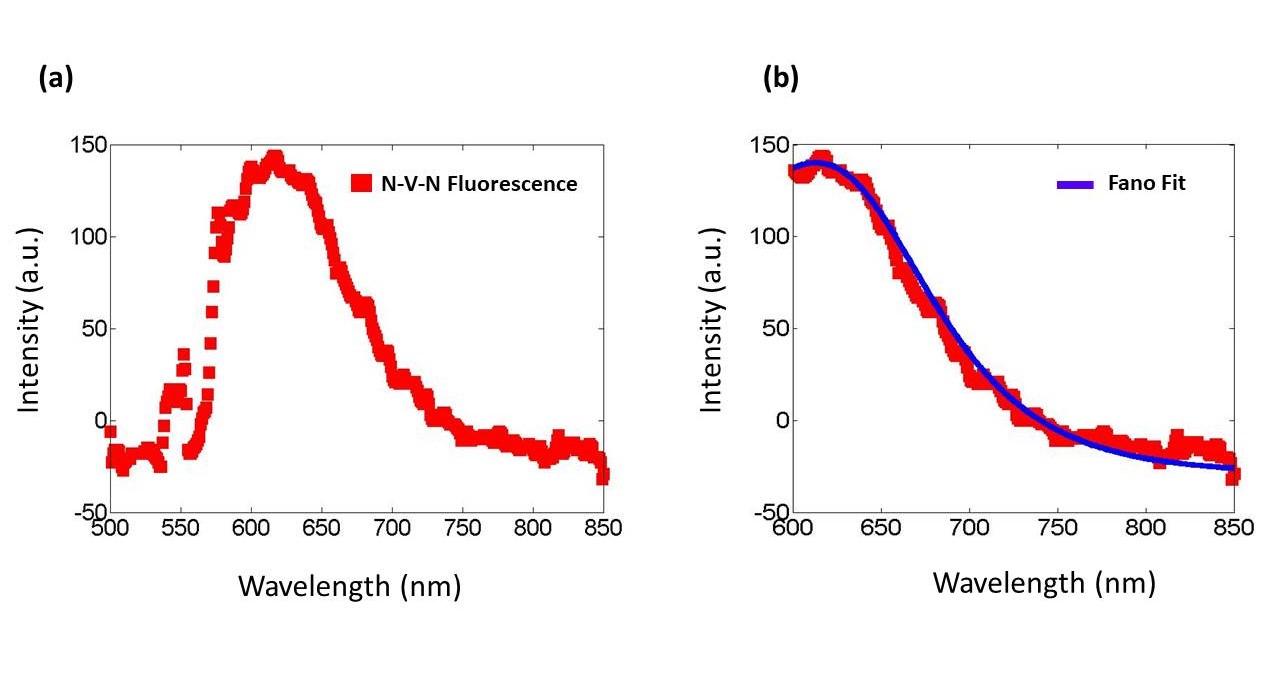}
\caption{(a) Red curve: Typical  fluorescence spectral trace in N-V-N solution displaying asymmetric lineshape. (b) Blue curve: Fano-like fit using Eq. 1. Numerical fit corresponds to a central wavelength $\lambda_0=650$ nm ($\omega_0=4.61 \times 10^{14}$ Hz), a linewidth $\gamma=1/4800  \times 10^{15}$ and a  Fano-like parameter $F=0.9$. }
\label{fig:stream}
\end{figure}


  Moreover, the optical properties of the N-V center are well suited for bio-imaging applications, with optical absorption in the range 490–560 nm and emission in the red - near infrared (637–800 nm), away from most fluorescent cell components. The emission occurs in a spectral window of low absorption attractive for biological labeling due to greater penetration of light in the surrounding tissue. A remarkable feature of N-V and N-V-N centers is that they do not photo-bleach or blink even under continuous high energy excitation conditions making them superior to conventional chromophores owing to their unprecedented photostability. A typical asymmetric fluorescence trace in N-V-N solution is displayed in Figure 1 (b).\\

In this paper, we present an analysis of photoluminiscence spectral traces in nanodiamond solutions in terms of Fano-like (i.e., asymmetric) lineshapes, were the total photoluminescence intensity can be considered the incoherent sum of the intensity due to two (or more) scattering and fluorescent processes, namely fluorescence due to nanodiamonds particle emission, and scattering due to the background solution. Therefore, in principle it should be possible to tune  the shape of the photoluminiscent spectra by modifying different relevant parameters, such as sample concentration or temperature, or by the addition of three or more scattering processes to the solution.  We are argue that, analogously to Fano resonace, the strong sensitivity of the asymmetric curve shape to different parameters can find relevant application in precision sensors, switches and filters, with the additional advantage  that nanodiamond fluorescence is highly bio-compatible and photo-estable,  features which can readily extend the applicability of Fano sensors, switches and filters to the biological domain.\\ 

\section*{Results}

The fluorescence measurement setup is depicted in Fig 2. A fiber diode laser (Thorlabs LP Series) was used as excitation source. Typical spectral traces characterizing the excitation laser are displayed in inset of Figure 2(a). The laser beam was reflected by a dichroic beam-splitter with a separation wavelength around 600 nm. The beam was then focused by a microscope objective lens (100X, NA = 1.3, Olympus) into a cuvette containing nanodiamond solutions placed between microscope sample holders. Fluorescence emission from the sample was collected with the same objective lens (Olympus x100) and passed through the dichroic beamsplitter. It was further filtered using a long-pass filter with a cut-off wavelength at 650 nm and focused into a fiber-coupled spectrometer (Thorlabs CCS series). Confocal micrsocope images were obtained by replacing the Spectrometer (Thorlabs CCS series) by a CCD camera (Thorlabs WFS series). Figure 2(b) displays a confocal microscope image of single  NV fluorescence. The fluorescence from single emitter as depicted in the Figure 2(b) is  roughly 250 brighter than the background. The registred photostability is of several minutes, only limited by the mechanical stability of the setup (see Supplementary Material \cite{31}). The confocal images were used for aligning the spectrometer and measure fluorescence in a single bright spot.\\

In Figure 3 (a) we present a typical spectral trace produced by N-V-N solutions. In Figure 3 (b) we present a Fano-fit to a typical spectral trace for fluorescence  in ND solution. The predicted  asymmetric fluorescence spectral trace can be described in terms of Fano-like lineshapes due to competing independent processes (Eq. 1). Fano resonance is a type of resonant scattering phenomenon that gives rise to an asymmetric line-shape due to interference between a background and a resonant scattering process which produces the asymmetric line-shape. Eventhough fluorescence in the time scales considered in this work is not a coherent phenomena, the total scattered intensity can be considered as the incoherent superposition of intensities arising from different scattering process, in particular scattering due to watery solution and due to nanodiamond particles, and fluorescence emission. The superposition of these two independent processes can give rise to an asymmetric spectrum as displayed in Figure 1 (b) (theory) and Figure 3 (a) (experiment). Moreover by tuning the relative concentration as well as scattering properties of the liquid and the particles it should be possible to tune the sharp peak characterizing the fluorescent spectrum. Thus all the outstanding metrology applications typically assigned to Fano resonances can also be applied to such nanodiamond solutions, with the additional benefit that nano-diamond fluorescence is highly compatible with bio-metric applications.\\

We fitted the experimental trace to a Fano-like curve using Equation  1. For a central wavelength $\lambda_0=650$ nm ($\omega_0=4.61 \times 10^{14}$ Hz) and a linewidth $\gamma=1/4800 \times 10^{15}$ Hz, we obtained a Fano-like factor $F=0.9$ (blue curve in Figure 4 (b)), thus confirming the high degree of asymmetry of the lineshape, and its ascribed unique metrological applications.

\section*{Discussion}

Fano-like spectral features  arise from the superposition between two (or more) scattering and fluorescence processes, for this reason they are expected to possess an inherent sensitivity to changes in geometry, composition, concentration or local environment: therefore small perturbations can induce dramatic lineshape shifts. This property renders Fano-like media particularly attractive for a range of applications, many not yet foreseen.

Perhaps the most straightforward application of Fano-like nanodiamond  solutions is in the development of tunable chemical or biological sensors. Here, the introduction of a solution of molecules in the immediate environment of a Fano-like nano-diamond solution can induce an unexpectedly large spectral shift in the fluorescent frequency. Detection limits that approach single-molecule binding events may be possible. This effect, in combination with biomarkers properties of nanodiamonds, could enable a new generation of label-free chemical and bioanalysis probes adaptable to high-throughput applications. 

Additionally, efficient sensing has
been demonstrated in asymmetric split-ring resonators, and in a dolmen-type structure, where the Electro-Magnetic-Induced Fano resonance shows a sensitivity to the presence of glucose in aqueous solution. In the latter case, the active sensing geometry of an asymmetric structure, with its localized light modes probing attolitre or zeptolitre volumes around each sensing element, is simple and mass-producible. This geometry possesses very high sensitivities of nominally 600 nm per refractive-index unit and figures of merit in the range of 5. This feature could be emulated in tailored nanodiamond solutions which, in combination with bio-compatibility of nanodiamond due to low blinking, low toxicity and emission far from typical biological spectral, can find several promising applications in glucose bio-sensing and bio-switching. 

Finally, a Fano-like nanodiamond fluorescence can find applications in detection of atomic single layers in biological tissues, with enhanced sensitivity to local displacement or deformation.

\section*{Acknowledgements}

The author is grateful to Friedemann Rienhardt, Philip Neumann, and Ilja Gerhardt for assistance in confocal microscope constuction and sample preparation. The author gratefully acknowledes insightfull discussions with Pablo Bianucci. This work was supported by PICT Startup 2015 0710, and UBACyT PDE 2017.

\section*{Author contributions statement}

GP conceived the idea, conducted experiments, analized the data, and wrote the manuscript. 

\section*{Additional information}

The are no competing interests.

\end{document}